\documentclass[myepj-spec,pdftex]{mySvjour}
\usepackage{graphicx}
\usepackage{hyperref}
\usepackage{color}
\usepackage{xspace}
\usepackage[square,sort&compress,numbers]{natbib}   
\usepackage{verbatim}
\usepackage{amsmath,amssymb,amsfonts} 
\usepackage{tabularx}
\usepackage{multicol}
\usepackage{footnote}
\usepackage{listings}
\usepackage[switch, modulo]{lineno}

\usepackage{lmodern,bm}                
\usepackage[T1]{sansmath} 
\SetMathAlphabet{\mathsfbf}{sans}{\sansmathencoding}{\sfdefault}{bx}{sl}
\usepackage{etoolbox}
\AtBeginEnvironment{sansmath}{}{}{}

\usepackage{lipsum}

\definecolor{darkblue1}{rgb}{0,0,.2}
\definecolor{darkblue}{rgb}{0,0,.2}
\definecolor{darkred}{rgb}{0.5,0,0}
\pagecolor{white} 
\color{black}     
\hypersetup{breaklinks=true, 
	colorlinks=true, 
	linkcolor=darkblue1, 
	menucolor=darkblue1, 
	urlcolor=darkblue1,
	citecolor=darkblue1,
	pdftitle={},
	pdfauthor={},
	pdfsubject={},
	pdfkeywords={},
	pdfproducer={}
}
\usepackage{siunitx}

\bibstyle{plain}
\newcommand{\bi}{\begin{itemize}}
\newcommand{\ei}{\end{itemize}}

\newcommand{\ben}{\begin{enumerate}}
\newcommand{\een}{\end{enumerate}} 

\newcommand{\bt}[1]{\begin{table}[tb]\begin{tabular}{#1} \hline\hline  \\[-1.0em]}
\newcommand{\et}[2]{\hline\hline \end{tabular} \caption{#1} \label{#2} \end{table}}

\newcommand{\be}{\begin{equation}}
\newcommand{\ee}{\end{equation}}

\newcommand{\bea}{\begin{eqnarray}}
\newcommand{\eea}{\end{eqnarray}}

\newcommand{\bc}{}

\newcommand{\mev}{\ensuremath{\mathrm{\,Me\kern -0.1em V}}\xspace}
\newcommand{\gev}{\ensuremath{\mathrm{\,Ge\kern -0.1em V}}\xspace}

\begin{document}
	
	\twocolumn[{%
		\begin{@twocolumnfalse}
			
			\begin{flushright}
				\normalsize
			\end{flushright}
			
			\vspace{-2cm}
			\title{\Large\boldmath Searching for Gravitational Waves with CMS}
			%

\author{Kristof Schmieden \inst{1} \and Matthias Schott \inst{1}}
\institute{\inst{1} Institute of Physics, Johannes Gutenberg University, Mainz, Germany}


			\abstract{
The idea of searching for gravitational waves using cavities in strong magnetic fields has recently received significant attention. 
Most concepts foresee moderate magnetic fields in rather small volumes, similar to those which are currently employed for axion-like particle searches. 
We propose to use the magnet system of the Compact Muon Solenoid (CMS) experiment after the high luminosity phase of the LHC as a key component for a future detector for gravitational waves in the MHz frequency range. 
In this paper we briefly discuss a possible cavity concept which can be integrated into CMS and additionally provide a first estimation of its possible sensitivity.}	
	\maketitle
	\end{@twocolumnfalse}
}]

\tableofcontents

\section{Introduction}	

With the first observation of gravitational waves (GWs) by the LIGO and Virgo interferometers \cite{LIGOScientific:2016aoc} a new era of astronomy has begun. Our current understanding of the universe implies the existence of gravitational waves with varying frequencies over several orders of magnitude, starting from super massive binary black hole systems in the nHz regime to kHz for compact binary objects and up to GHz for GWs from the cosmic gravitational microwave background \cite{Aggarwal:2020olq}.

The experimental concept of interferometers has been proven highly successful and new generations, such as the future Einstein Telescope \cite{Punturo:2010zz} are planned. One alternative concept is based on the usage of electromagnetic cavities, either being pumped (e.g. \cite{Bernard:1999uz, Bernard:2002ci}) or being placed in a magnetic field. The latter approach has been discussed in more detail most recently \cite{Berlin:2021txa, Domcke:2022rgu, Ringwald:2020ist}, as similar experimental setups are currently performed or planned in the context of searches for axion-like particles, e.g. \cite{ADMX:2021nhd, Schmieden:2021msb,Berlin:2022hfx}. The basic physics principle can be simply understood: a gravitational wave will alter the shape of the cavity, thus inducing a change in the magnetic flux through the cavity, which in turn generates an electric signal which can be detected. Clearly this depends on the frequency and the incoming direction of the gravitational wave as well as the resonance frequencies of the cavity itself. The sensitivity to gravitational waves using a cavity-based experiment has been derived in \cite{Berlin:2021txa} and can be summarised by the signal power
\begin{equation}\label{eq:signalPower}
P_{sig} = \frac{1}{2} Q \omega_g^3 V^{5/3} (\eta _n h_0 B_0)^2 \frac{1}{\mu_0 c^2},
\end{equation}
with $\omega_g$ denoting the GW frequency and $h_0$ the magnitude of the GW strain. The cavity is described by its volume $V$, its quality factor $Q$ as well as the external magnetic field $B_0$. The dimensionless coupling constant $\eta _n$ is given by  
\begin{equation}
\eta_n = \frac{|\int_{V} d^3x E^*_n \cdot \hat j_{+,\times}|}{V^{1/2} (\int _V d^3x |E_n|^2)^{1/2}},
\end{equation}
where $E_n$ denotes a resonant mode of the cavity and $\hat j_{+,\times}$ describe spatially-dependent dimensionless functions for the spatial profile and polarization of the GW signal. We refer to \cite{Berlin:2021txa} for further details. It is important to note that the sensitivity increases quadratically with the magnetic field strength $B_0$ and to the power of 5/3 with the cavity volume V.

In the context of axion searches, each axion mass corresponds to exactly one resonance frequency. Searching over a certain axion mass range therefore requires varying resonance frequencies of the cavity. Hence typical cavity experiments, such as ADMX, rely on a tunable cavity design, or are directly designed for a broadband search, such as the future MadMax experiment \cite{Caldwell:2016dcw}. The situation is different for GW searches, as their signal is not expected to be localised at one frequency but is instead expected to be broadband. 

Most axion cavity search experiments focus on the axion mass range between $0.5\cdot10^{-3}$ and $10^{-7}$ eV, since axion models in this parameter space would be candidates for dark matter and could potentially solve the strong CP problem. 
This mass range corresponds to resonance frequencies from $25\,\mathrm{MHz}$ to $100\,\mathrm{GHz}$, i.e. cavity dimensions from a few cubic centimeters to a few cubic meters. The size of the cavity is therefore mainly determined by the axion mass that is searched for. The situation is different for GW detection applications as standard cosmological sources of GWs imply frequencies between $10^{-4}\mathrm{Hz}$ and $10^{10}\mathrm{Hz}$, requiring therefore significantly larger cavity volumes.

One of the largest magnetic volumes in terms of $\sim V^{5/3} B_0^2$ is available at the Compact Muon Solenoid (CMS) experiment \cite{CMS:2008xjf} at the Large Hadron Collider (LHC). After the high luminosity phase of the LHC in mid 2040, the central detector system could in principle be removed and replaced by a dedicated GW cavity experiment. In the following we briefly introduce the CMS magnet system (Section \ref{sec:CMS}), before discussing a potential concept for the GW cavity (Section \ref{sec:Cavity}). The expected sensitivity is summarized in Section \ref{sec:Sensitivity}.    

\section{The Compact Muon Solenoid \label{sec:CMS}} 	

The magnet system of the CMS experiment \cite{CMS:1997moj} consists of a NbTi superconducting solenoid and a magnetic flux return yoke made of construction steel. The solenoid creates a magnetic flux of 130 Wb in a cylindrical volume with a diameter of nearly 6m and a length of 12.5~m. A central magnetic flux density of 3.8 T is created by an operational direct current of 18.164 kA. 
It is installed in a vacuum cryostat in the central barrel wheel of the magnet yoke. The magnet yoke includes two three-layer barrel wheels around the solenoid cryostat on each side of the central barrel wheel as well as  two nose disks inside each end of the solenoid cryostat, and four endcap disks on each side of the cryostat \cite{Klyukhin:2021ldf}. The muon detection system of the CMS experiment is partly integrated within the return yoke, while the main parts of the remaining sub-detector systems, i.e. the tracking detectors as well as the electromagnetic and hadronic calorimeters, are within the solenoidal magnetic field. Those detector components would not be needed for a cavity-based search for gravitational waves and have to be removed to provide space for the GW cavity.

\section{Cavity Concept\label{sec:Cavity}}

The cavity design is largely determined by the potential background sources of the final experiment, namely external electromagnetic signals, mechanical vibrations, thermal radiation as well as potential effects from cosmic rays.

The cavity will be placed inside a cryostat and cooled to 2K or even lower to reduce the thermal radiation. 
The cavity will act by itself as a Faraday cage and hence largely shield any external electromagnetic backgrounds. The cryostat will act as an additional shielding cylinder as it surrounds the actual GW cavity and is specifically designed to also block external electromagnetic signals from entering the detection volume.

While not being strictly necessary, an operation of the GW cavity in vacuum might be beneficial, otherwise the interaction of cosmic rays with the gas volume could lead to ionization electrons which orbit in the magnetic field of the cavity\footnote{The effect of cosmic rays on the cavity performance could be studied insitu as the muon tracking system of the CMS experiment would still be in place.}, or pressure differences in gas might lead to mechanical oscillations. 

Mechanical oscillations of the GW cavity can be reduced by hanging the cavity as a nearly free floating object within the magnetic field. Remaining internal oscillations and deformations can be tracked with laser control systems on the outside walls of the GW cavity. 

\begin{figure*}[thb]
\centering
	\resizebox{0.49\textwidth}{!}{\includegraphics{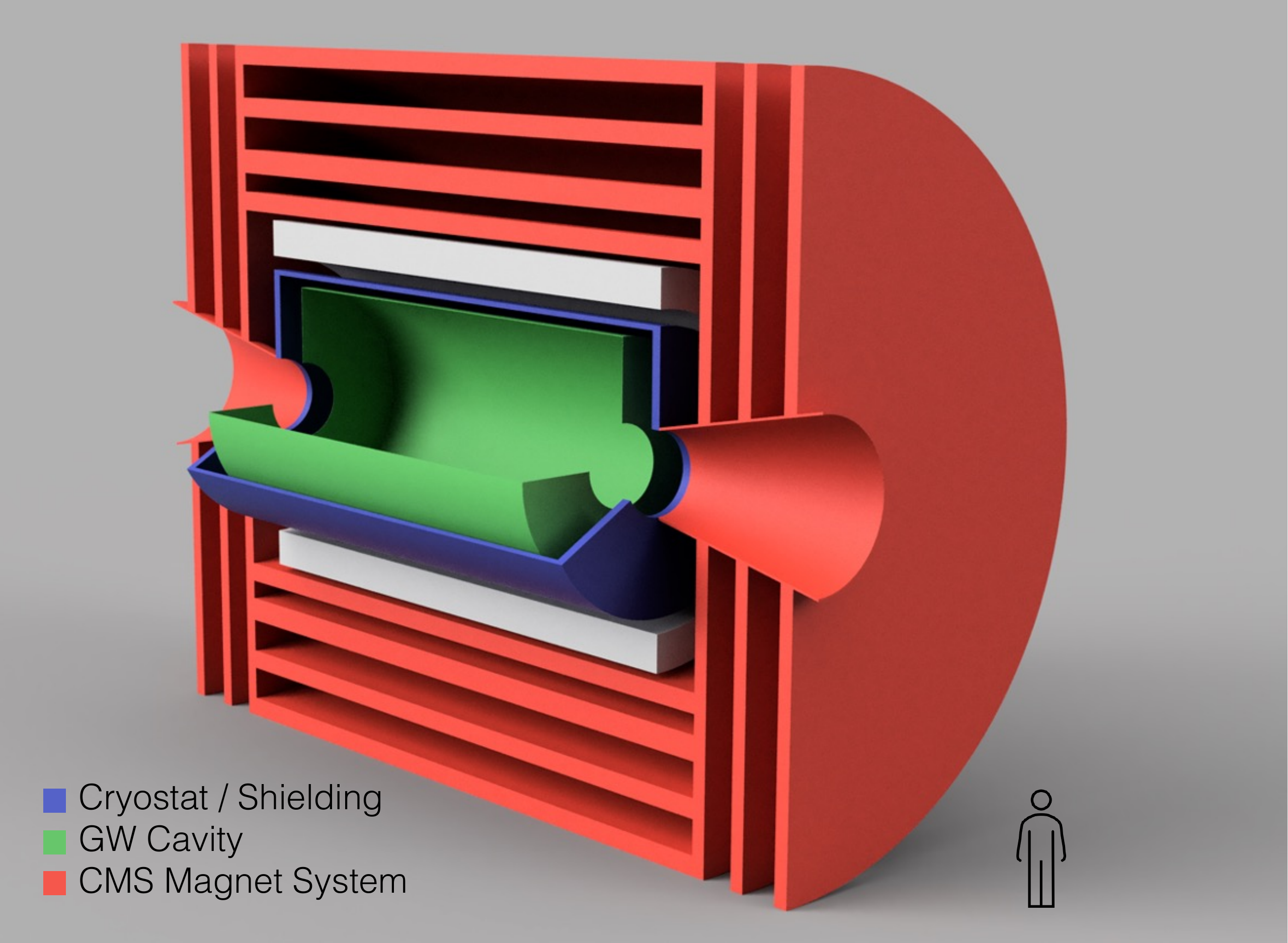}}
	\resizebox{0.49\textwidth}{!}{\includegraphics{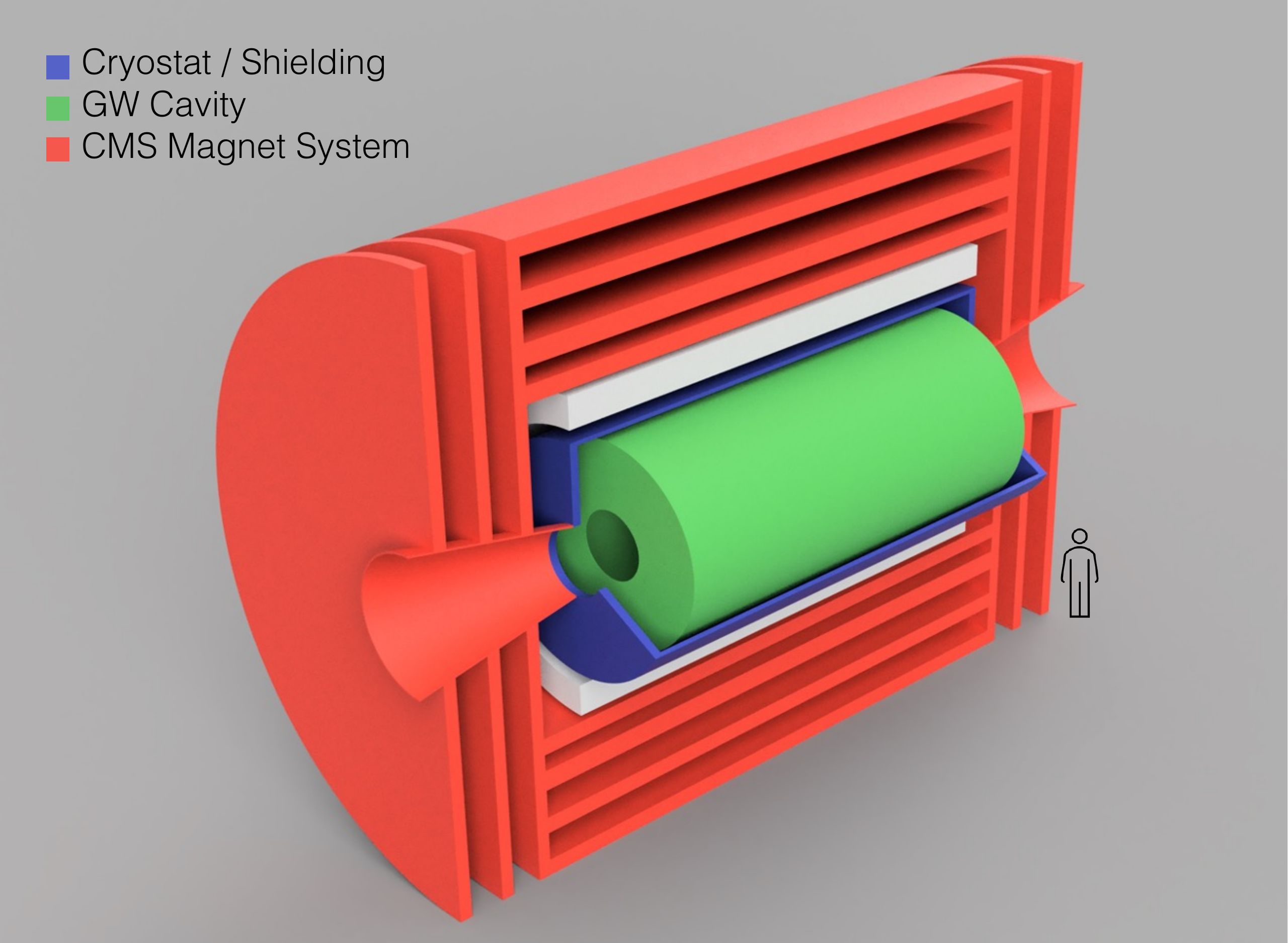}}
\caption{Schematic illustration of the CMS magnet system, the shielding cylinder as well as the gravitational wave cavity.}
\label{fig:Cavity}
\end{figure*}

These considerations suggest the following setup: a cryostat cylinder with an inner radius of 2600\,mm, an inner length of 11800\,mm, and a thickness of 200\,mm could be installed directly within the CMS magnetic volume. The actual GW cavity with a radius of 2400\,mm and length of 10000\,mm, can be constructed rather thin and therefore light, allowing it to be mounted on free hanging devices within the shielding cylinder. The dimensions have been chosen such to ensure a very homogeneous magnetic field of 3.8T with deviations smaller than 5\%. The vacuum would be created within the full volume of the shielding cylinder, reducing significantly the mechanical stress on the GW cylinder that would arise if the vacuum was instead created only there. A schematic of the full setup is shown in Figure \ref{fig:Cavity}.

\section{Expected Sensitivity \label{sec:Sensitivity}}
\begin{figure*}[thb]
\centering
	\resizebox{0.49\textwidth}{!}{\includegraphics{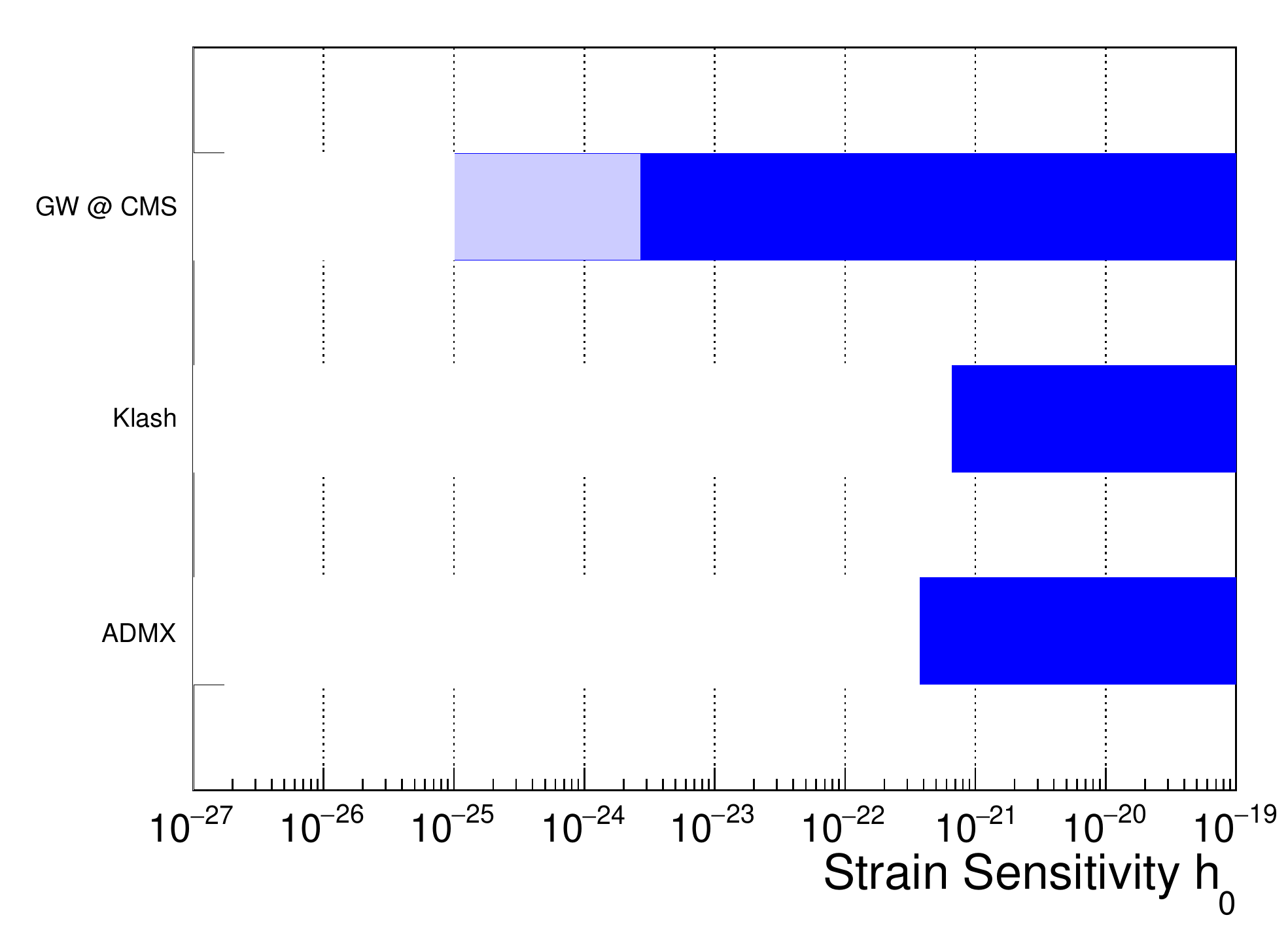}}
	\resizebox{0.49\textwidth}{!}{\includegraphics{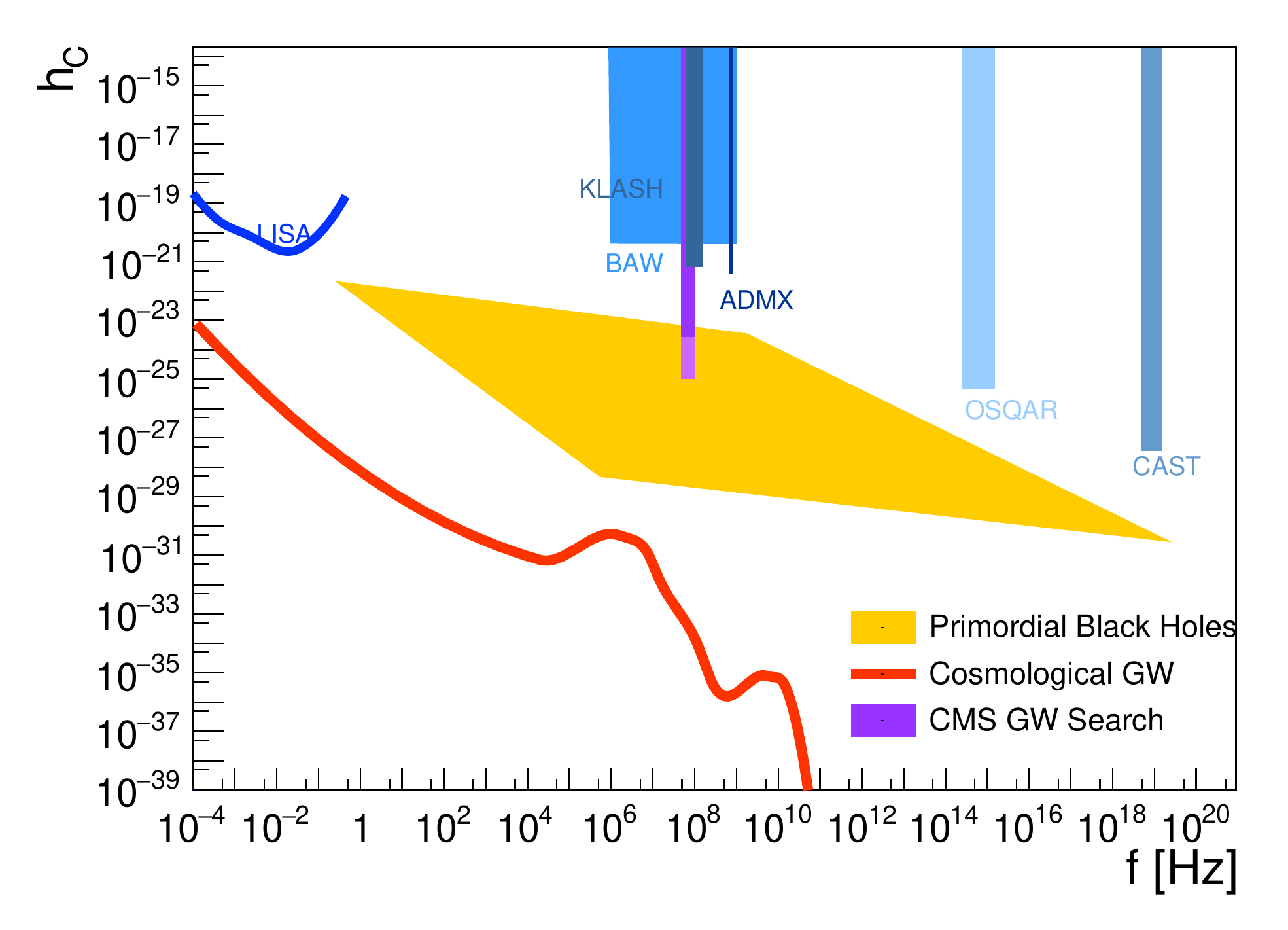}}
\caption{Sensitivity of a GW search using the CMS magnet system. Left: Expected sensitivity of the GW strain for the proposed experiment (option 2 shaded, option 1 solid) in comparison to the sensitivity of the ADMX \cite{ADMX:2021nhd} and Klash proposal \cite{Alesini:2019nzq} experiments.
Also shown are the sensitivity of Bulk Accustik Wave devices (BAW) \cite{Goryachev:2014yra,Page:2020zbr,Goryachev:2021zzn} as well as the reinterpretations of the results from OSQAR \cite{OSQAR:2015qdv} and CAST \cite{CAST:2017uph} measurements presented in \cite{Ejlli:2019bqj}. Right: Expected GW spectrum from cosmological sources \cite{Ringwald:2022xif} and primordial black holes \cite{Franciolini:2022htd} in comparison to the sensitivity of various experiments, including this proposal. The range of PBH mergers shown in yellow covers masses from $10^{-6} M_\odot$ to $10^{-16} M_\odot$ with decreasing strain and time before merging from 1 Gyr to $10^{-3}$~sec with increasing frequency.}
\label{fig:Sensitivity}
\end{figure*}

Using the signal power as defined in eq. \ref{eq:signalPower} the signal-to-noise ratio (SNR) can be calculated using the Dicke radiometer equation as
\begin{equation}
    \texttt{SNR} \approx \frac{P_{sig}}{k_B T_{sys}}\sqrt{\frac{t_{int}}{\Delta\nu}}, 
\end{equation}
where  $T_{sys}$ is the effective noise temperature, $t_{int}$ is the measurement integration time, and 
$\Delta\nu$ is the signal frequency bandwidth.
Requiring $\mathrm{SNR} > 1$, in line with previous calculations, e.g. \cite{Berlin:2021txa}, the strain sensitivity is calculated as
\begin{equation}
    h_0 \ge \left(\frac{1}{\omega_g}\right)^{3/2} \left(\frac{1}{V } \right)^{5/6} \sqrt{\frac{k_B T_{sys}}{Q}}\left(\frac{\Delta\nu}{t_{int}}\right)^{1/4} \frac{\sqrt{2\mu_0} c}{\eta_n B_0}.
\end{equation}
Using the cavity parameters as listed in tab. \ref{tab:cavity} we expect a strain sensitivity of the proposed setup of $h_0 > 2.7\cdot 10^{-24}$ ($1\cdot 10^{-25}$) for the two listed options. This is two orders of magnitude better than the sensitivity of ADMX or Klash \cite{Alesini:2019nzq}, where the latter would operate in a similar frequency range. 
The scaling of the sensitivity with the volume of the cavity is mostly canceled by the decreasing frequency with increasing cavity radius as $f_0 \propto 1/r$. The sensitivity to $h_0$ scales with the radius $r$ and length $l$ of the cavity as $h_0 \propto 1/(r^{1/6} \cdot l^{5/6})$. 
Option 1 and 2 in tab. \ref{tab:cavity} differ in their assumptions on the quality factor, the temperature of the cavity, as well as integration time which is one day and one year respectively. 
For $\Delta\nu$ we assume $180\,\mathrm{Hz}$, which is about three times the cavity line width, as smaller values become increasing impractical in this frequency regime. The same value is used for a comparison to Klash, while for ADMX we use $\Delta\nu = 10\,\mathrm{kHz}$, which is close to the line-width of the ADMX cavity. 
While experiments searching for axions aim to scan a large frequency range in a short period of time and have typical integration times on the order of  1\,min, searches for a GW continuum can rely on few frequency points with large integration times in the order of days. 
Changing the integration time from 2 min to one day only changes the sensitivity by a factor of 5.   
A detailed comparison of the sensitivity of the proposed setup to existing experiments is shown in tab. \ref{tab:cavity} and Figure.  \ref{fig:Sensitivity}. The sensitivity reaches into the range of primordial black hole (PBH) mergers with masses down to $10^{-9} M_\odot$ and signal times in the order of days, as predicted by \cite{Franciolini:2022htd}. 

\begin{table}[h]
    \centering
    \begin{tabular}{c|c|c|c|c}
         & \multicolumn{2}{c|}{GW @ CMS} & ADMX & Klash\\
        & opt. 1& opt. 2& & \\
        \hline\hline
        $f_0$[MHz] & 48 & 48  & 650 & 62 \\
        V [$m^3$] & 181 & 181 & 0.136 & 22.2\\
        Q & $10^6$ & $2\cdot 10^7$ & $8\cdot 10^4$ & $7\cdot 10^5$\\
        B [T] & 4 & 4 & 7.5 & 0.6\\
        $\Delta\nu$ [Hz] & 180& 180 & $10^4$ & 180\\
        $T_{sys}$[K] & 2 & 1 & 0.6 & 4\\
        $t_{int}$ [s] & $10^5$& $3\cdot 10^7$ & 120 & $120$ \\
        \hline
        $h_0 \ge $ & $2.7\cdot 10^{-24}$ & $1\cdot 10^{-25}$ & $3.7\cdot 10^{-22}$ & $6.5\cdot 10^{-22}$
    \end{tabular}
    \caption{Comparison of the cavity parameters and sensitivity to the gravitational strain $h_0$ of this proposal to ADMX and the Klash \cite{Alesini:2019nzq} proposal.}
    \label{tab:cavity}
\end{table}

\section{Conclusion}

In this paper we propose to reuse the magnet system of the CMS experiment  after the high luminosity phase of the LHC for a dedicated cavity-based search for gravitational waves. The expected sensitivity on $h_0$ after one year of running would be $6\cdot 10^{-25}$ on $h_0$. 

We are aware that this is a rather futuristic endeavour with tremendous technical challenges. However, there are two more decades before the end of the High-Luminosity LHC. This time can be used to develop several smaller prototype experimental setups using existing magnet systems of previous experiments, such as DZero \cite{D0:2005cnn} or CDF \cite{CDF:1988lbl}. Given the outstanding physics case, we believe that it might be worthwhile to investigate this scenario further.


\section*{Acknowledgement}
We thank Pedro Schwaller for the helpful discussions and comments during the preparation of this paper, as well as Jack MacDonald for the proof-reading. We also thank Tim Schneemann for providing the graphical illustration of the cavity experiment. This work would have not been possible without the ERC-Grant ``LightAtTheLHC'' as well as the continuous support from the PRISMA+ Cluster of Excellence at the University of Mainz.  
	
\bibliographystyle{apsrev4-1} 
\bibliography{./Bibliography}
	
\end{document}